\documentclass[twocolumn,prb,aps]{revtex4}
\usepackage{graphics}
\begin{document}
\title{Electron-phonon interaction in the three-band model}
\author{O. R\"osch  and O. Gunnarsson}
\affiliation{ Max-Planck-Institut f\"ur Festk\"orperforschung,
Postfach 800665, D-70506 Stuttgart, Germany}

\begin{abstract}
We study the half-breathing phonon in the three-band model
of a high temperature superconductor, allowing for vibrations of
atoms and resulting changes of hopping parameters. Two different
approaches are compared. From the three-band model a $t$-$J$ model
with phonons can be derived, and phonon properties can be calculated. To
make contact to density functional calculations, we also study the
three-band model in the Hartree-Fock (HF) approximation. The
paramagnetic HF solution,
appropriate for the doped cuprates, has similarities to the
local-density approximation (LDA). However, in contrast to the LDA,
the existence of an antiferromagnetic insulating solution for the
undoped system makes it possible to study the softening of the
half-breathing phonon under doping. We find that although the HF
approximation and the $t$-$J$ model give similar softenings, these
softenings happen in quite different ways. We also find that the HF
approximation gives an incorrect doping and ${\bf q}$ dependence for
the softening and too small a width for the (half-)breathing phonon.

\end{abstract}
\maketitle
\section{Introduction}
The electron-phonon coupling for high-$T_c$ cuprates has recently
attracted much interest. Lanzara {\it et al.}\cite{Lanzara} discovered
strong coupling to a mode at 70 meV in many cuprates. The coupling
was ascribed to a half-breathing phonon along the (1,0,0) direction.
This is an in-plane bond-stretching mode, where the vibrations are
primarily due to two of the four O atoms surrounding a Cu atom in
the CuO$_2$ plane. Lanzara {\it et al.}\cite{Lanzara} deduced a rather
strong apparent electron-phonon coupling $\lambda \sim 1$. The
half-breathing phonon shows an  anomalous softening when the cuprates
are doped, in particular towards the zone
boundary.\cite{Pintschovius1,Pintschovius2,Pintschovius3,anomol}
The softening of other phonons upon doping can be explained as a
screening of the ions in the doped system.\cite{Pintschovius1} The softening
of the half-breathing mode, however, cannot be described in a shell model
with conventional parameters.\cite{Pintschovius1} This supports the
idea that this phonon has substantial electron-phonon coupling, which would
lead to a reduction of the frequency in the doped but not in the undoped
system. This phonon has an appreciable broadening,\cite{Pintschovius2}
which can also be explained in terms of a substantial electron-phonon
coupling. From the broadening one could estimate an electron phonon
coupling for this mode of the order of $\lambda \sim 0.2-0.3$, using
the formula of Allen\cite{Allen} and the density of states of 
Mattheiss.\cite{Mattheiss} Anomalous 
behavior of bond-stretching modes has also been observed in other 
compounds.\cite{Tranquada,Braden,Reichardt}

The half-breathing phonon has been studied\cite{Becker,Horsch,Nagaosa,Oliver}
within the $t$-$J$ model,\cite{Zhang} and a substantial softening was
found.\cite{Becker,Horsch,Oliver} Phonons and the electron-phonon interaction 
have also been studied\cite{Pickett,Savrasov,Bohnen} extensively within the 
local-density approximation (LDA).\cite{KS} Bohnen {\it et al.}\cite{Bohnen}
found phonon frequencies in good agreement with experiment for YBa$_2$Cu$_3$O$_7$. 
In particular, the frequency of the half-breathing mode along the (1,0,0) 
direction was found to be anomalously soft, in agreement with experiment. 
Since LDA does not describe antiferromagnetism in the undoped system, however, 
it cannot properly describe phonons in the undoped case. It is therefore not
clear how much the phonon is softened under doping in the LDA. Furthermore, 
LDA calculations show a weak electron-phonon coupling to the half-breathing
phonon, with $\lambda$ at the zone boundary being $\sim$ 
0.01.\cite{Bohnen} This is in disagreement with the large 
width of the half-breathing phonon, which is believed to be due to
a rather strong electron-phonon coupling. The weak coupling raises questions
about the reason for the low frequency of the half-breathing phonon in 
the LDA calculation for the doped system. 

To address these issues, we here study the three-band model,\cite{Emery}
including a Cu $d_{x^2-y^2}$ and two O $p$-orbitals per unit cell.
We allow displacements of the atoms from their equilibrium positions
and take the corresponding changes of hopping integrals into account. In
this model we can study phonons.  We solve the model using a Hartree-Fock (HF)
mean-field approximation, which may be expected to simulate features
of the LDA for the doped system. For instance, we find a similar width
for the half-breathing phonon as in the LDA. In contrast to LDA, however,
this approximation gives an antiferromagnetic solution for the undoped
system. We can therefore obtain the softening upon doping within
this framework. The softening of the half-breathing phonon is indeed
found to be of the same order of magnitude as the experimental result, 
supporting the idea that the LDA can describe the softening.

Alternatively, a $t$-$J$ model with phonons can be derived from the
three-band model. This model is solved using exact diagonalization.\cite{Oliver}
Thereby, many-body effects are included, and the interplay between
electron-phonon and electron-electron interactions are considered 
in a more explicit way. In this paper we compare the results in these 
two approaches.

We find that the HF solution of the three-band model and the $t$-$J$
model give a comparable softening of the half-breathing phonon.
This happens, however, in a very different way in the two approaches.
Furthermore, we find that the dependence on doping and on ${\bf q}$
is rather different in the two approaches. In the $t$-$J$ model, the
softening is $\sim \delta$ for small $\delta$, while the doping dependence 
is weaker in the HF solution. The $t$-$J$ model gives a smaller softening
for the ${\bf q}=(1,1)\pi/a$ breathing mode than for the ${\bf q}=
(1,0)\pi/a$ half-breathing mode, while the opposite is found in the HF
approximation. In both cases $t$-$J$ results are in better agreement
with experiment.

In Sec. II we introduce the models and in Sec. III the methods. The
results are presented in Sec. IV. The results of the HF approximation
for the three-band model are compared with the exact diagonalization
results of the $t$-$J$ model in Sec. V.

\section{Models}
We study the three-band model with $N$ unit cells
\begin{eqnarray}\label{eq:2.1}
H&=&\varepsilon_d \sum_{i\sigma} n_{id\sigma}
+\varepsilon_p\sum_{j\sigma}n_{jp\sigma}
+U\sum_i n_{id\uparrow}n_{id\downarrow} \\
&&+\sum_{\langle i,j\rangle\sigma}\lbrack
t_{ij}^{pd}\psi^{\dagger}_{id\sigma}
\psi_{jp\sigma}^{\phantom \dagger}+h.c.\rbrack
+\sum_{\langle i,j\rangle\sigma}t_{ij}^{pp}
\psi^{\dagger}_{ip\sigma}
\psi_{jp\sigma}^{\phantom \dagger}\nonumber,
\end{eqnarray}
where $n_{id\sigma}$ and $n_{jp\sigma}$ are the occupation numbers for
the $N$ $3d$-orbitals and the $2N$ $2p-$orbitals, respectively,
and $\psi_{id\sigma}$ and $\psi_{jp\sigma}$ are the corresponding
annihilation operators. The site energies are $\varepsilon_d$ and
$\varepsilon_p$. There is a Coulomb integral $U$, acting between
two $3d$-electrons on the same Cu site.  The model includes hopping between
nearest neighbor Cu and O atoms and between the O atoms which are
nearest neighbors of a particular Cu atom.\cite{t} These hopping 
integrals are given by $t^{pd}$ and $t^{pp}$, respectively, with the 
signs determined by the relative orientations of the orbitals involved.
Displacing an atom leads to changes of both electrostatic potentials
and hopping integrals. Electrostatic potentials are screened differently
in the doped and undoped systems. It was found, however, that the change
of screening does not strongly influence the half-breathing 
phonon.\cite{Pintschovius2} We therefore assume here that the 
changes of hopping integrals are the dominating effects.\cite{Falter}

We assume the hopping integral $t^{pd}$ has a power dependence on
the atomic separation $d$, i.e.,
\begin{equation}\label{eq:2}
t^{pd}(d)=\left({a/2\over d}\right)^{3.5}t^{pd0}
\end{equation}
where $a$ is the lattice parameter. The exponent 3.5 for $t^{pd}$
was estimated from LDA band structure calculations.\cite{Ove}
We neglect the phonon modulation of $t_{pp}$.

This model can be solved directly as it stands within some 
approximation, here the Hartree-Fock approximation. We can, 
however, also derive a $t$-$J$ model with phonons, and then 
solve this model. The $t$-$J$ model has one site per Cu atom. 
In the undoped system, there is one hole per site. Doping introduces
additional holes, which primarily sit on O sites. These O holes form
Zhang-Rice singlets with neighboring Cu holes.\cite{Zhang}
It is then assumed that each site is occupied by either a $d$-hole
or a Zhang-Rice singlet.  Expressing the atomic displacements in terms
of phonon operators and working to linear order in the displacements
 leads to the $t$-$J$ model with phonons\cite{Oliver}
\begin{eqnarray}\label{eq:2.2}
H_{t\textrm{-}J}&=&
J\sum_{<i,j>}\left(
{\bf S}_i\cdot{\bf S}_j-\frac{n_in_j}{4}
\right)
-t\sum_{<i,j>\sigma}\tilde c_{i\sigma}^{\dagger}\tilde c_{j\sigma}^{\phantom\dagger}\\
&&+\sum_{{\bf q}\nu}\omega_{\nu}({\bf q})\left(b_{{\bf q}\nu}^{\dagger}
b^{\phantom\dagger}_{{\bf q}\nu} +{1\over 2}\right)\nonumber \\
&&+\sum_{ij\sigma}\tilde c_{i\sigma}^{\dagger}\tilde c_{j\sigma}^{\phantom\dagger}
\sum_{{\bf q}\nu}g^{t\textrm{-}J}_{ij}({\bf q},\nu)(b_{{\bf q}\nu}^{\phantom\dagger}+
b_{-{\bf q}\nu}^{\dagger})\nonumber.
\end{eqnarray}
Here $\tilde c_{i\sigma}^{\dagger}$ creates a $d$-hole on site $i$ if this
site previously had no hole, i.e., if it had a Zhang-Rice singlet.     
The operator $b_{{\bf q},\nu}^{\dagger}$ creates a
phonon with wave vector ${\bf q}$, index $\nu$ and frequency $\omega_{\nu}
({\bf q})$. $\delta_{i,j-\hat x}=1$ if site $j$ is one site to the right
of $i$ in the $x$-direction. The formulas for $g_{ij}^{t\textrm{-}J}({\bf q},\nu)$
have been given elsewhere.\cite{Oliver} While we treat the three-band model
in the electron picture, we treat the $t$-$J$ model in the hole picture.

In deriving this model, we assumed that $t_{pd}\ll
|\varepsilon_d-\varepsilon_p|$, $U$. This is a rather poor approximation
for realistic parameters. Here we have somewhat artificially reduced the 
value of $t^{pd0}$ to 1.2 eV, which gives a realistic\cite{Prelovsek} 
value of $t=-0.47$ eV if perturbation theory is used, and probably also 
more realistic values for $g_{ij}^{t\textrm{-}J}$, by compensating for
some of the effects of perturbation theory.\cite{Oliver} The treatment 
above neglected quadratic terms, although these also can give 
a doping dependent contribution to the phonon softening.

\section{Methods}

We use a frozen phonon method to calculate the phonon frequency
in the HF approximation for the three-band model. This gives
the effective Hamiltonian
\begin{eqnarray}\label{eq:3.0}
H_{\rm{eff}}&=&\sum_{i\sigma} (\varepsilon_d +U\langle n_{id-\sigma}\rangle)
n_{id\sigma} +\varepsilon_p\sum_{j\sigma}n_{jp\sigma} \\
&&-U\sum_i \langle n_{id\uparrow}\rangle\langle n_{id\downarrow}\rangle \nonumber\\
&&+\sum_{\langle i,j\rangle\sigma}\lbrack
t^{pd}_{ij}\psi^{\dagger}_{id\sigma}
\psi_{jp\sigma}^{\phantom \dagger}+h.c.\rbrack
+\sum_{\langle i,j\rangle\sigma}t^{pp}_{ij}
\psi^{\dagger}_{ip\sigma}
\psi_{jp\sigma}^{\phantom \dagger}\nonumber,
\end{eqnarray}
where $\langle n_{id-\sigma}\rangle$ is the expectation value of
$n_{id-\sigma}$. We also introduce the effective level
\begin{equation}\label{eq:3.0a}
\varepsilon_{d\sigma}^{\rm eff}=\varepsilon_d +U\langle n_{id-\sigma}\rangle.
\end{equation}

\begin{figure}[t]
\centerline{
{\rotatebox{0}{\resizebox{7.7cm}{!}{\includegraphics {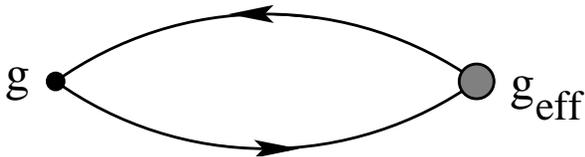}}}}}
\caption[]{\label{fig:1}Diagram describing the linear response.
$g \sim \partial H/\partial u$ and $g_{\rm{eff}} \sim \partial
H_{\rm{eff}}/\partial u$ describe the bare and screened perturbations
due to a phonon.  }
\end{figure}

A static distortion is built into the lattice and the change of the
total energy is calculated. From this the interaction with the electrons
can be deduced and the softening of the phonon obtained. The displacements
of the atoms in the lattice is denoted by $\lbrace u\rbrace$. The energy
of the electronic system is then
\begin{equation}\label{eq:3.1}
E\lbrack \lbrace u\rbrace \rbrack={\rm Tr}
(H\lbrack \lbrace u\rbrace \rbrack
\rho\lbrack \lbrace u\rbrace \rbrack),
\end{equation}
where $\rho\lbrack \lbrace u\rbrace \rbrack$ is the density matrix,
obtained in some approximation, e.g. the HF approximation.
Using the Hellmann-Feynman theorem, it follows that
\begin{equation}\label{eq:3.2}
{\partial^2 E \over \partial u^2}= {\rm Tr} \left(
{\partial H \over \partial u}
{\partial \rho \over \partial u}
\right)+{\rm Tr}\left( {\partial ^2 H\over \partial u^2}\rho\right)
\equiv {\partial^2 E \over \partial u^2}^{(1)}+
{\partial^2 E \over \partial u^2}^{(2)}.
\end{equation}
From the total energy calculation, we can deduce both $\partial^2 E /
\partial u^2$ and $(\partial^2 E /     \partial u^2)^{(i)}$, $i$=1, 2.
$(\partial^2 E /\partial u^2)^{(1)}$ describes how a first order
change in the external Hamiltonian leads to a first order change in
the density matrix, which acts back at the Hamiltonian. This contribution
can be calculated in linear response
\begin{equation}\label{eq:3.3}
{\partial \rho\over \partial u}=\chi {\partial H_{\rm{eff}}\over \partial u},
\end{equation}
where $\chi$ is the response function for noninteracting electrons
and the derivative of the effective Hamiltonian is obtained from
the total energy calculation. Thus we have
\begin{equation}\label{eq:3.4}
{\partial^2 E \over \partial u^2}^{(1)}\equiv {\rm Tr}
\left( {\partial H \over \partial u}{\partial \rho\over
\partial u}\right)= {\rm Tr} \left( {\partial H\over \partial u}\chi
{\partial H_{\rm{eff}}\over \partial u}\right).
\end{equation}
We introduce a dielectric function
\begin{equation}\label{eq:3.5}
\epsilon={\rm Tr} \left({\partial H\over \partial u} \chi {\partial H\over \partial u}\right)
/     {\rm Tr} \left({\partial H\over \partial u}\chi {\partial H_{\rm{eff}}\over \partial u}\right).
\end{equation}
The result in Eq. (\ref{eq:3.4}) can also be obtained from a diagrammatic
technique.  It is described by a bubble diagram, as shown in Fig. \ref{fig:1}.
This is the only diagram which enters in the HF approximation. We observe
that the screening of the perturbation, described by $g_{\rm{eff}}$, enters at
one of the vertices. Using $g_{\rm{eff}}$ at both vertices would lead to
double counting. By using the diagrammatic approach, we can obtain not only
a contribution to the energy but also the width of the phonon.

\section{Results}

Below we present results for the three-band model in the Hartree-Fock
approximation, using  the parameters $t^{pd0}=1.6$ eV, $t^{pp}=0$ and
$U=8$ eV. The lattice parameter is $a=3.8$ \AA. We have adjusted
$\varepsilon_p$ so that the separation between the effective $3d$-level
and $2p$-levels
\begin{equation}\label{eq:5.1}
\varepsilon_d^{\rm eff}-\varepsilon_p^{\rm eff}\equiv
\varepsilon_d+U\langle n_{d\sigma}\rangle-\varepsilon_p=3 \ {\rm eV},
\end{equation}
where $\langle n_{d\sigma}\rangle$ is the average occupation of the
$3d$-levels per spin. This is a typical LDA estimate for $\varepsilon_d^{\rm eff}
-\varepsilon_p^{\rm eff}$ in  a three-band model.\cite{t}
We perform two calculations, one for the
undistorted lattice and one for a lattice where a phonon has been
built in. This gives the second derivative, $\partial^2 E/\partial u^2$.
From this derivative we obtain the softening of the phonon due to the
interaction with the electrons in the model, reducing the frequency
$\omega_{ph0}$ to $\omega_{ph}$. Here $\omega_{ph0}$ is assumed to be due
to forces not included in the model in Eq.~(\ref{eq:2.1}), e.g.,
electrostatic forces and core-core overlap effects. The calculations
were performed for a cluster of $32\times 32$ CuO$_2$ units and periodic
boundary conditions. The doping was chosen in such a way that degenerate
levels were either completely full or completely empty, i.e., all ``shells''
were either full or empty.

\begin{figure}[t]
\centerline{
{\rotatebox{-90}{\resizebox{6.0cm}{!}{\includegraphics {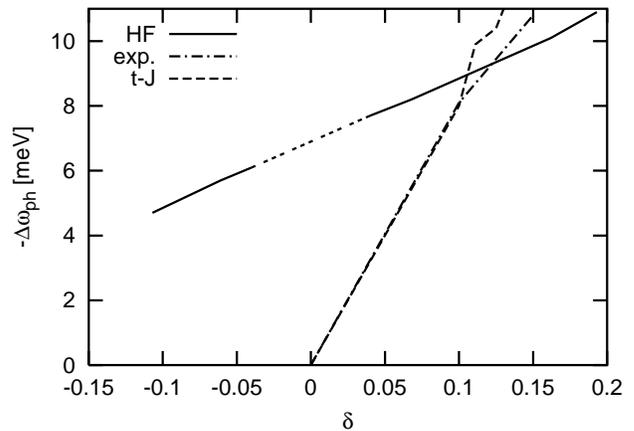}}}}}
\caption{\label{fig:2}Softening $\Delta \omega_{ph}$ for the zone boundary
half-breathing phonon. Results are given in the HF approximation (full line),
the $t$-$J$ model (dashed line) and according to experiment (dash-dotted
line) as a function of the hole doping $\delta$. The lines serve as
a guide for the eye between the few points in the $t$-$J$ model and
according to experiment. The HF approximation refers to results for
the shift in a paramagnetic calculation for doping $\delta$ relative
to an antiferromagnetic calculation for $\delta=0$. The dashed part
of the HF line indicates schematically that the systems goes
antiferromagnetic for small dopings.}
\end{figure}

We first consider the half-breathing phonon for ${\bf q}=(1,0)\pi/a$,
where the two O atoms at ($a$/2,0) and \mbox{(-$a$/2,0)} move towards (or away from)
from the Cu atom at (0,0). We perform a calculation for the undoped system,
having five electrons per unit cell, and allowing for spin-polarization.
We adjust $\omega_{ph0}$ so that the softened frequency $\omega_{ph}$ is
0.080 eV for the zone boundary half-breathing phonon, as is found experimentally.

The result as a function of the hole doping $\delta$ ($\delta<0$
means electron doping) is shown in Fig. \ref{fig:2}. The doping
dependence is relatively weak. Since the undoped system is antiferromagnetic
but the doped system is (assumed to be) paramagnetic, as found experimentally,
the softening is not necessarily small for small dopings. The spin-polarized
system has a large gap of about 4.6 eV. Due to this gap, the response of
the system to a phonon is substantially weaker than for the paramagnetic
state. For instance, $\lbrack \partial^2 E /\partial u^2\rbrack^{(1)}$ 
contributes a softening of only about 4 meV for the spin-polarized system 
but 6 meV for a non spin-polarized system with a similar number of electrons. 
Similarly, the contributions of $\lbrack \partial^2 E /\partial u^2\rbrack^{(2)}$ 
are about 8 and 13 meV, respectively.

We find that
screening reduces the quantity $\lbrack \partial^2 E /\partial u^2\rbrack^{(1)}$
by about a factor of two, i.e., $\epsilon\approx 2$ in Eq. (\ref{eq:3.5}).
By evaluating the diagram in Fig. \ref{fig:1}, we find that the zone boundary
half-breathing phonon is broadened by about 0.4 meV for $\delta=0.16$. This is 
similar to what Bohnen {\it et al.}\cite{Bohnen} found in an LDA calculation.

It is interesting to compare the half-breathing phonon and the ${\bf q}=
(1,1)\pi/a$ breathing phonon. In the latter case all four O atoms surrounding
a Cu atom move towards this atom. In the model considered above ($t^{pp}=0$)
and for $\delta=0$, the model has nesting for ${\bf q}=(1,1)\pi/a$. The result
would be a very strong response for small $\delta$, making the calculation
rather meaningless. Instead we consider $t^{pp}=1.1$ eV.\cite{t} This changes
the Fermi surface and avoids an unrealisticly strong nesting. In this case 
we adjust $\omega_{ph0}$ so that $\omega_{ph}=0.090$ eV in the spin-polarized 
calculation for the undoped system, as found experimentally.  
For the doping $\delta=0.096$ we find that the ${\bf q}=(1,1)\pi/a$
breathing phonon is softened by 11 meV. This is a larger softening 
than was found in Fig. \ref{fig:2} for this doping. We have also performed
a calculation for the ${\bf q}=(1,0)\pi/a$ half-breathing phonon in this
model and for this doping. The softening is 7 meV, which again is smaller 
than for the the breathing phonon. This is contrary to experiment, where a
larger softening is found for the half-breathing phonon.

\section{Comparison of the the $t$-$J$ model and HF solution
of the three-band model}

To make contact with the $t$-$J$ model, we transform the three-band model
to a one-band model. We assume that the O $2p$-levels are far below the
Cu $3d$-levels (in the electron picture). We can then project out the 
O $2p$-levels and obtain a model with just effective Cu $3d$-levels. 
This model can be compared with the $t$-$J$ model, since both models
have one band. Here we focus on the linear response term
$(\partial^2 E / \partial u^2)^{(1)}$ in Eq.~(\ref{eq:3.4}),
which are of the same order of magnitude in the $t$-$J$ model and
the HF approximation for a typical $\delta\sim 0.1$. 

We introduce a projection operator $P$, which projects out the Cu $3d$-levels,
and its complement $Q=1-P$. Considering the resolvent operator of the HF
Hamiltonian, we obtain
\begin{eqnarray}\label{eq:4.1}
&&P (z-H)^{-1}P \\
&&\quad= \lbrack P(z-H)P-PHQ(z-QHQ)^{-1}QPH\rbrack ^{-1},\nonumber
\end{eqnarray}
where $z$ is some typical energy.
We then obtain the effective one-band parameters
\begin{eqnarray}\label{eq:4.2}
&&\varepsilon_{id\sigma}^{\rm One}=\varepsilon_d
+\sum_{\langle j\rangle}{(t_{ij}^{pd})^2\over z-\varepsilon_p}, \nonumber \\
&&t_{ii^{'}}^{\rm One}={t_{ij^{\phantom '}}^{pd}t_{ji^{'}}^{pd}\over z-
\varepsilon_p},
\end{eqnarray}
where the sum over $j$ for $\varepsilon_{id\sigma}^{\rm One}$ runs over
the nearest O neighbors of Cu and $j$ for $t_{ii^{'}}^{\rm One}$ refers to
the common nearest neighbor O atom of the Cu atoms $i$ and $i^{'}$.
We first consider the unperturbed (no phonon) system. We choose $z$ to 
be in the middle of the band and solve the self-consistent equations
\begin{eqnarray}\label{eq:4.3}
&&\varepsilon_d^{\rm{One}}=\varepsilon_d
+{4(t^{pd})^2\over \varepsilon_d^{\rm{One}}-\varepsilon_p}\\
&&t^{\rm One}_{ii^{'}}=-{(t^{pd})^2\over \varepsilon_d^{\rm{One}}-
\varepsilon_p},  \nonumber
\end{eqnarray}
where the factor 4 for $\varepsilon_d^{\rm{One}}$ comes from the four 
O neighbors of a Cu atom.

We then introduce a phonon in the system and ask for the linear response
of the electronic system. A perturbation term is introduced in the 
one-band Hamiltonian
\begin{equation}\label{eq:4.4}
\partial \varepsilon_{id\sigma}=\sum_{\langle j\rangle}
2{t_{ij}^{pd}\partial t_{ij}^{pd}\over \varepsilon_d^{\rm{One}}-\varepsilon_p},
\end{equation}
where $\partial t_{ij}^{pd}$ are the changes in the Cu-O hopping integrals.
To linear order there is no change in $t_{ii^{'}}^{\rm One}$.
 
For a half-breathing phonon at the zone boundary, the on-site perturbation is
\begin{equation}\label{eq:4.5}
{\partial H \over \partial u}u=4t_{pd}{\partial t_{pd}\over \partial r}
{1\over \varepsilon_d^{\rm{One}}-\varepsilon_p}u,
\end{equation}
where $u$ is the absolute value of the phonon amplitude. The quantity 
$\partial H/ \partial u$ enters in the calculation of $(\partial^2 E/
\partial u^2)^{(1)}$ in Eq.~(\ref{eq:3.4}).

In the $t$-$J$ model, the on-site perturbation is\cite{Oliver}
\begin{equation}\label{eq:4.6}
4t_{pd}{\partial t_{pd}\over \partial r}\left({2\lambda^2-1\over
|\varepsilon_d-\varepsilon_p|}
+{2\lambda^2 \over U-|\varepsilon_d-\varepsilon_p|}\right)u,
\end{equation}
where $\lambda=0.96$. The first term comes from the hopping of a $3d$-hole
into the O $2p$-states and the second term from the hopping of a O $2p$-hole
into the Cu $3d$-state. The second term has no correspondence in
Eq.~(\ref{eq:4.5}). Equation~(\ref{eq:4.6}) has an additional factor 2 coming
from a phase coherence factor in the Zhang-Rice singlet. This results
from the singlet being explicitely written as a sum of two terms. Both these
effects are genuine many-body effects. The $-1$ in the first term in
Eq.~(\ref{eq:4.6}) results from taking the difference in the energy gain of a
Zhang-Rice singlet and a single $3d$-hole.

\begin{table}[bt]
\caption[]{\label{table:4.1}Contributions to the phonon softening in
the $t$-$J$ model and in the HF solution of the three-band model.
Here $\Delta\equiv |\varepsilon_d^{\rm{One}}
-\varepsilon_p|= |\varepsilon_d-\varepsilon_p|$.
}
\begin{tabular}{llll}
\hline
\hline
  Source & $t$-$J$ & HF &Ratio \\
\hline
Coupling  &$\lbrack (2\lambda^2-1)/\Delta
+2\lambda^2/( U-\Delta)\rbrack^2$ &$(1/ \Delta)^2$ & $\approx 3$  \\
Sum rule & $\approx 2\delta\pi N $ & $\approx \pi N$ & $\approx 2 \delta$ \\
Screening & 1  & $\approx 0.5 $ & $\approx 2$   \\
Denominator &  &  & $\approx 1$ \\
\hline
Product  &     &  &      12$\delta$ \\
\hline
\hline
\end{tabular}
\end{table}

We find that this on-site coupling dominates over the off-site coupling.\cite{Oliver}
In the qualitative discussion we therefore only consider the on-site coupling
$g_{ii}^{t\textrm{-}J}({\bf q}) \equiv g({\bf q)}$. The phonon self-energy in the $t$-$J$ model is
\begin{equation}\label{eq:2.8a}
\Pi({\bf q},\omega)=
{g({\bf q})^2\chi^{t\textrm{-}J}({\bf q},\omega)\over 1+g({\bf q})^2
\chi^{t\textrm{-}J}({\bf q},\omega)D_0({\bf q},\omega)},
\end{equation}
where $D_0({\bf q},\omega)$ is the noninteracting phonon Green's function.
We find that the second term in the denominator is small for the parameters
considered here. In the formal discussions below we therefore neglect it. The
phonon self-energy is then proportional to the response function. This result 
can then be directly compared with the result in Eq.~(\ref{eq:3.4}) for the
one-band model. In both cases the response function is multiplied by 
the appropriate coupling constant squared, given by Eq.~(\ref{eq:4.5})
and Eq.~(\ref{eq:4.6}), respectively.

To compare the $t$-$J$ model with the one-band model, we put $|\varepsilon_d^{\rm{One}}
-\varepsilon_p|= |\varepsilon_d-\varepsilon_p|\equiv \Delta=3$ eV and $U=8$
 eV.\cite{cons} We find that the square of Eq.~(\ref{eq:4.6}) is about 
a factor of three larger than the square of Eq.~(\ref{eq:4.5}). This difference 
due to differences in coupling constants is shown in Table \ref{table:4.1}.

The linear response of the one-band model is given by
\begin{eqnarray}\label{eq:4.7}
&&\chi^{\rm HF}({\bf q},\omega)=2\sum_{\bf k} \lbrack 1-f({\bf k}+{\bf q})\rbrack
f({\bf k}) \\
&&\quad\times \left( {1 \over \omega - \omega({\bf k},{\bf q}) +i0^+}
-  {1 \over \omega + \omega({\bf k},{\bf q}) -i0^+}\right)  \nonumber
\end{eqnarray}
where $f({\bf k})$ is the Fermi function for a state with wave vector
${\bf k}$ and energy $\varepsilon({\bf k})$ and $\omega({\bf k},{\bf q})=
\varepsilon({\bf k}+{\bf q})-\varepsilon({\bf k})$. We have the sum-rule
\begin{equation}\label{eq:4.8}
{1\over N}\sum_{{\bf q}\ne 0}\int_{-\infty}^{\infty}\left| {\rm Im}\chi^{\rm HF}
({\bf q}, \omega+i0^{+})\right|=4\pi n(1-n)N\approx \pi N,
\end{equation}
where $N$ is the number of sites and $n$ is the fractional filling
 of the band. Typically, we are interested in a system with 
$n=(1-\delta)/2 \approx 0.5$ electrons per site and spin, which leads 
to the right hand side of Eq.~(\ref{eq:4.8}). For the $t$-$J$ model 
we ask for the carrier-carrier response function
\begin{eqnarray}\label{eq:4.9}
\chi^{t\textrm{-}J}({\bf q},\omega)&=&\sum_{\nu}|\langle \nu|\rho({\bf q})|0\rangle|^2 \\
&&\times \left({1\over \omega-\omega_{\nu}+i0^{+}}-{1\over \omega+\omega_{\nu}-i0^{+}}
\right), \nonumber
\end{eqnarray}
where $|\nu\rangle$ is an excited many-electron state with the 
excitation energy $\omega_{\nu}$ and $\rho({\bf q})$ is carrier 
density operator. One finds\cite{Horsch}
\begin{equation}\label{eq:4.10}
{1\over N}\sum_{{\bf q}\ne 0}\int_{-\infty}^{\infty}\left| {\rm Im}\chi^{t\textrm{-}J}({\bf q},
\omega+i0^{+})\right|=2\pi N \delta(1-\delta).
\end{equation}
Eqs.~(\ref{eq:4.8}) and (\ref{eq:4.10}) differ by approximately a factor of
$2\delta$ for small $\delta$ (see Table \ref{table:4.1}).

The two approaches further differ by the screening in the HF approach,
discussed in Eq.~(\ref{eq:3.5}). This reduces the HF result by roughly
a factor of two. Finally, we have to consider that a typical energy
denominator in Eq.~(\ref{eq:4.8}) and Eq. (\ref{eq:4.9}) are different.
Calculating the average $\langle 1/\omega \rangle$, we find comparable
results for the two models. As can be seen from Table \ref{table:4.1},
the difference between the two approaches results in a ratio of about $12\delta$. 
For $\delta \sim 0.1$, the two results are then similar. This indicates why
the $t$-$J$ model and the HF solution of the three-band model can give similar
softening of the half-breathing mode although the physics is quite different.

We next consider the imaginary part of the phonon self-energy $\Pi$, which gives
the phonon width. As an orientation, we first consider a simple model.
Since Im $\chi(\omega)\sim \omega$ for small $\omega$, we assume
\begin{equation}\label{eq:4.11}
{\rm Im}\ \Pi({\bf q},\omega)=\cases{A\omega,&if $|\omega|\le W$;\cr
0,&otherwise,\cr}
\end{equation}
where $A$ is some constant. From the Kramers-Kronig relation, we can then
derive
\begin{equation}\label{eq:4.12}
{\gamma\over -\Delta \omega_{ph}}=\pi{\omega_{ph}\over W},
\end{equation}
where $\gamma=2{\rm Im} \ \Pi({\bf q},\omega)$ is the full width at half
maximum of the phonon and $\Delta \omega_{ph}$ is its shift.

\begin{figure}[t]
\centerline{
{\rotatebox{-90}{\resizebox{6.0cm}{!}{\includegraphics {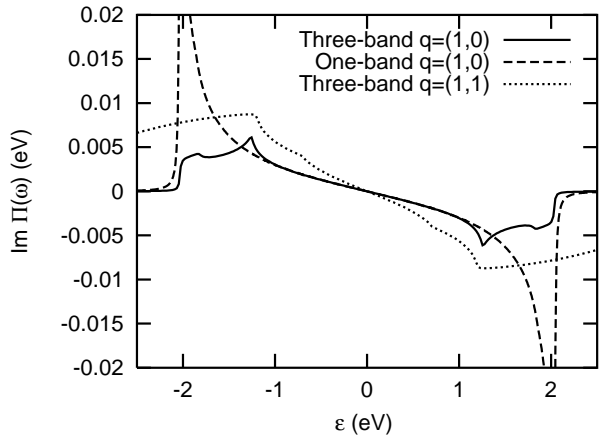}}}}}
\caption[]{\label{fig:4.1}Im $\Pi({\bf q},\omega)$ for the
${\bf q}=(1,0)\pi/a$ half-breathing mode in the three-band model (full line)
and the one-band model (dashed line) and for the ${\bf q}=(1,1)\pi/a$
breathing mode (dotted line). The doping is $\delta=0.1$.
}
\end{figure}

Figure \ref{fig:4.1}
compares Im $\Pi({\bf k},\omega)$ for the half-breathing phonon in the
one- and three-band models and the breathing phonon in the three-band model.
The one-band model was constructed to describe what happens close to $E_F$
and therefore the one- and three-band models agree very well for small $\omega$.
For $|\omega|\gg 0$, the one-band model gives a larger $|{\rm Im}\ \Pi({\bf q},
\omega)|$ than the three-band model, and $|{\rm Re}\ \Pi({\bf q},\omega)|$
is overestimated correspondingly in the one-band model. Appropriate numbers
for the the half-breathing phonon and the one-band model are $W=2$ eV,
$\omega_{ph}=0.07$ eV, $\Delta \omega_{ph}=-0.007$ eV. Inserting this in
Eq.~(\ref{eq:4.12}) leads to $\gamma=0.8$ meV. This is about twice the width
actually calculated. The reason for this overestimate is that Im $\Pi(
{\bf q},\omega)$ is actually smaller than assumed in Eq.~(\ref{eq:4.10})
for small $\omega$.

Figure \ref{fig:4.1} also shows the HF result  for the ${\bf q}=
(1,1)\pi/a$ breathing phonon. These results were obtained for $t^{pp}=1.1$
eV. There is then no strong nesting of the Fermi surface. The HF approximation,
nevertheless, gives a larger broadening for the ${\bf q}=(1,1)\pi/a$
breathing phonon than for the ${\bf q}=(1,0)\pi/a$ half-breathing phonon,
whether the latter is calculated for $t^{pp}=0$ or $t^{pp}=1.1$ eV. As is
well-known,\cite{Pickett} this is due to the fact that the wave vector
${\bf q}=(1,1)\pi/a$ fits better to the traces of nesting left over for
$t^{pp}=1.1$ eV.

\begin{figure}[t]
\centerline{
{\rotatebox{-90}{\resizebox{6.0cm}{!}{\includegraphics {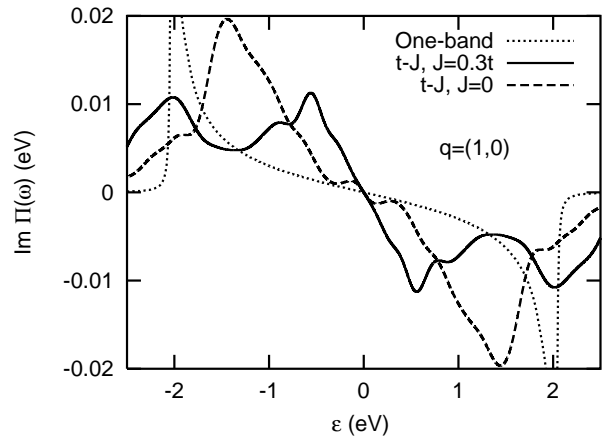}}}}}
\caption[]{\label{fig:4.2}Im $\Pi({\bf q},\omega)$ for the
${\bf q}=(1,0)\pi/a$ half-breathing mode in the $t$-$J$ model for $J/t=0.3$
(full line) and $J/t=0$ (dashed line) and in the one-band model (dotted
line). The results for the $t$-$J$ model were obtained for a $4\times 4$ cluster.
The doping is $\delta=0.125$.}
\end{figure}

To study the phonon width in the $t$-$J$ model, we calculate the phonon
self-energy.\cite{self} We calculate the phonon spectral function $B({\bf q},\omega)$
using exact diagonalization. A Hilbert transform can then be used to obtain
the phonon Green's function $D({\bf q},\omega)$. The phonon  self-energy
$\Pi({\bf q},\omega)$ is calculated by inverting
\begin{equation}\label{eq:2.11}
D^{-1}({\bf q},\omega)={D_0^{-1}({\bf q},\omega)-\Pi({\bf q},\omega)},
\end{equation}
where $D_0({\bf q},\omega)$ is the noninteracting phonon Green's function.
This approach has important advantages for small systems. $B({\bf q},\omega)$
has too few structure to determine the phonon width. A broadened version of
$\Pi({\bf q},\omega)$, however, can give such information.\cite{self}
Results for $\Pi({\bf q},\omega)$ are shown in Fig. \ref{fig:4.2}.

In the view of Eq.~(\ref{eq:4.12}), one might have expected the width
of the half-breathing phonon to be similar in the HF approximation
and the $t$-$J$ model, since the shifts are similar. This is not
true, however, since the frequency dependence differs strongly from the
linear dependence assumed in Eq.~(\ref{eq:4.11}). This is illustrated in
Fig. \ref{fig:4.2}. The figure shows that for the $t$-$J$ model
with $J=0$, some spectral weight has been moved to small frequencies
due to the hopping constraint, which creates low energy excitations.
This is even more true for the finite $J$ case ($J/t=0.3$). The present
clusters are too small to give reliable results, in particular for
the low-lying excitations, and the results above should be considered
as qualitative. They illustrate the general trend, however, of
transferring some spectral weight to small frequencies, which tends
to lead to a substantially increased width of the half-breathing
phonon.

\section{Summary}

We have studied the properties of the half-breathing phonon in the
three-band model of a high-$T_c$ cuprate. The results in a $t$-$J$
model with phonons and the HF approximation of the three-band model
were compared. Although the two approaches give similar softenings
at typical dopings $\delta$, the underlying physics is quite different.
The hopping constraint in the $t$-$J$ model, resulting from the strong
Coulomb repulsion, leads to a strong reduction of the response to a
phonon. This reduction is, however, partly compensated by several other 
effects.  In particular, the coherent hopping in the formation of the 
Zhang-Rice singlet tends to increase the coupling in the $t$-$J$ model.
We find that the doping and ${\bf q}$ dependences for the (half-)breathing
mode are quite different in the two approaches, with the $t$-$J$ model
giving better agreement with experiment. The HF approximation also
gives a too small width for the half-breathing phonon. We therefore 
conclude that many-body effects play an important role for the 
interactions of the half-breathing phonon with the electrons.

We thank O. Jepsen and K.-P. Bohnen for providing unpublished results
and O.K. Andersen, K.-P. Bohnen, P. Horsch, O. Jepsen, N. Nagaosa and
Z.-X. Shen for many useful discussions.


\begin{thebibliography}{*}
\bibitem{cons}Since we use the electron picture for the three-band model
and the hole picture for the $t$-$J$ model, the use of $|\varepsilon_d^{\rm eff}
-\varepsilon_p|=3$ eV in both models together with $U=8$ eV could be
considered inconsistent, but is in agreement with parameters typically used in
the two models.
\end{thebibliography}
\end{document}